\documentclass[a4paper, amsfonts, amssymb, amsmath, reprint, showkeys, nofootinbib, twoside,superscriptaddress]{revtex4-2}
\usepackage[english]{babel}
\usepackage{here}
\usepackage[utf8]{inputenc}
\usepackage{comment}
\usepackage{bm}
\usepackage{braket}
\usepackage{url}
\usepackage[colorinlistoftodos, color=green!40,prependcaption]{todonotes}
\usepackage[pdftex, pdftitle={Article}, pdfauthor={Author}]{hyperref}

\newcommand{\Tr}[1]{\mathrm{Tr}\Bigl[{#1}\Bigr]}
\newcommand{\TrL}[1]{\mathrm{Tr}\Bigl[{#1}}
\newcommand{\ReB}[1]{\mathrm{Re}\Bigl[{#1}\Bigr]}
\newcommand{\ReBL}[1]{\mathrm{Re}\Bigl[{#1}}
\newcommand{\bk}{\bm{k}}

\newcommand{\trg}{\int \mathrm{Tr}(g)dk_xdk_y}
\newcommand{\detg}{\int \sqrt{\mathrm{det}(g)}dk_xdk_y}

\begin{document}
\title{Quantum metric on the Brillouin Zone in correlated electron systems and its relation to topology for Chern insulators}
\author{Takahiro Kashihara}
   
\affiliation{Department of Physics, Kyoto University, Kyoto 606-8502, Japan}
\author{Yoshihiro Michishita}
\affiliation{RIKEN Center for Emergent Matter Science (CEMS), Wako, Saitama 351-0198, Japan}
\author{Robert Peters}
    \email[Correspondence email address: ]{peters@scphys.kyoto-u.ac.jp}
\affiliation{Department of Physics, Kyoto University, Kyoto 606-8502, Japan}
    \date{\today} 
\begin{abstract}
Geometric aspects of physics play a crucial role in modern condensed matter physics. The quantum metric is one of these geometric quantities which defines the distance on a parameter space  and contributes to various physical phenomena, such as superconductivity and nonlinear conductivity. 
Despite its importance, the quantum metric in interacting systems is poorly understood.
In this paper, we introduce a generalized quantum metric(GQM) on the Brillouin zone for correlated electron systems. This quantum metric is based on the optical conductivity that is written by single-particle Green's functions. We analytically prove that this definition is equivalent to the existing definition of the quantum metric in noninteracting systems and that it is positive semi-definite as necessary for a metric.
Furthermore, we 
point out the relationship between the GQM and the Chern number in interacting systems. 
We then numerically confirm these properties of the GQM in the Qi-Wu-Zhang model with and without interaction. 
We believe that the GQM will be a step toward generalizing the quantum metric to the interacting regime.
\end{abstract}
\maketitle
\section{Introduction}

In various areas of physics, geometric aspects play an essential role in understanding the fundamental structure, such as in the gauge theory or the topological classification of defects\cite{nakahara2018geometry}. 
In condensed matter physics, the geometry of eigenstates in the momentum space has attracted much attention in recent years~\cite{bernevig2013topological,PhysRevX.9.041015,doi:10.1126/sciadv.1501524,Tokura2018,PhysRevLett.115.216806}. One of the most important geometric quantities in momentum space is the Berry curvature. The integral of the Berry curvature over the momentum space is quantized, and the value is proportional to the Chern number, which contributes to the quantum Hall effect~\cite{PhysRevLett.49.405}. 

The counterpart to the Berry curvature is the quantum metric. While the imaginary part of the quantum geometric tensor is proportional to the Berry curvature, the real part of the tensor is the quantum metric~\cite{shapere1989geometric}.   The quantum metric is a metric that defines the distance on the momentum space. In recent studies, it has been revealed that the quantum metric is equipped with rich mathematical properties~\cite{https://doi.org/10.48550/arxiv.2205.15353,PhysRevB.104.045104,PhysRevB.106.165133} and plays a crucial role in various fields of condensed matter physics such as superconductivity~\cite{Peotta2015,PhysRevLett.126.027002,PhysRevLett.128.087002,2022arXiv220900007H}, stability of the fractional quantum Hall states~\cite{PhysRevB.90.165139,PhysRevLett.127.246403}, electrical  conductivity~\cite{PhysRevX.10.041041,PhysRevB.105.085154,PhysRevB.102.165151,Ahn2022-pc,Resta2011, PhysRevB.62.1666, Raffaele_Resta_2002,PhysRevB.104.134312,PhysRevResearch.4.013217,2021arXiv211107782H}, orbital magnetism~\cite{PhysRevB.94.134423}, nonadiabatic quantum evolution\cite{PhysRevLett.121.020401}, energy shift in excitons~\cite{PhysRevLett.115.166802}, maximally localized Wannier function~\cite{PhysRevB.56.12847}, and non-Hermitian physics~\cite{PhysRevLett.127.107402,PhysRevB.103.125302}.

The quantum metric has also been used as an important tool to characterize quantum phase transitions since the pioneering works Ref.~\cite{PhysRevE.74.031123} and Ref.~\cite{PhysRevLett.99.100603}. Although the integral of the quantum metric is not quantized, the relationship between the quantum metric and topological invariants is also actively investigated in recent studies. For example, inequalities between integrals of the quantum metric (concretely, the trace of the quantum metric and the quantum volume defined by the quantum metric) and the Chern number have been noted~\cite{Peotta2015,PhysRevB.104.045103, PhysRevB.103.L241102,Zhang_2022}. It was numerically confirmed that the Chern number is inferred by the quantum volume~\cite{PhysRevB.104.045103}. Furthermore, the relation to fragile topology~\cite{PhysRevLett.124.167002,PhysRevLett.126.027002} and the winding number of the SSH chain (see the supplemental material of Ref.~\cite{PhysRevLett.124.167002}) have also been pointed out, and the general relationship between the quantum volume and the winding number of a chiral symmetric system, the Chern number, including the second Chern number, are investigated in detail in Ref.~\cite{10.21468/SciPostPhys.12.1.018}.

The topology of systems 
has been intensively studied even in correlated systems~\cite{Rachel_2018,Hohenadler_2013}. One of the strategies employed is to describe the topological invariant by the single-particle Green's function\cite{PhysRevLett.105.256803,PhysRevX.2.031008,article,PhysRevB.92.085126}. 
Since Green's functions can be defined even in the presence of correlations, topological invariants such as the TKNN invariant can be defined in correlated systems. It has been shown that these topological invariants are also quantized at zero temperature~\cite{PhysRevLett.105.256803,PhysRevB.83.085426}. The single-particle Green's function describes the quasi-particle states in the system, and the Chern number defined by the Green's function can be experimentally detected via the integer quantum Hall conductivity in correlated materials.

On the other hand, the quantum metric in interacting systems is not well understood. A very recent proposal of a definition of the quantum metric on the momentum space in interacting systems is given in Ref.~\cite{10.21468/SciPostPhysCore.5.3.040}. The authors propose a generalization of the quantum metric using the relationship between the quantum metric and the charge polarization susceptibility. In Ref.~\cite{10.21468/SciPostPhysCore.5.3.040}, the quantum metric is called "the dressed quantum metric." They define the dressed quantum metric using the Bloch wave function and the two-particles Green's function. 
Although they analyze the dressed quantum metric with impurity scattering by approximating the two-particles Green's function, the behavior of the dressed quantum metric in a general interacting system and its expression by the single-particle state are not investigated.
Furthermore, they do not mention the positive semi-definiteness of the dressed quantum metric in interacting systems, which must be satisfied as a metric. 
Thus, the quantum metric on the momentum space in interacting systems is still veiled.

Note that the quantum metric on the twist angle space can be defined even in interacting systems by using the ground state wave function \cite{Resta2011, PhysRevB.62.1666, Raffaele_Resta_2002}. However,  it is essentially different from the conventional definition of the quantum metric on the momentum space\cite{PhysRevB.56.12847} because it cannot be described in single-particle states.

In this paper, we define a generalized quantum metric(GQM) 
on the momentum space in interacting systems based on the single-particle Green's function without using Bloch wave functions. The GQM can be expressed by the eigenvalues and the eigenvectors of the single-particle spectral function. Thus, the GQM reflects the properties of the single-particle states in general systems. Generalizing the quantum metric on the momentum space using the single-particle spectral function will be a helpful tool in analyzing material properties because the quantum metric should affect physical properties even in interacting systems in the same way as in noninteracting systems~\cite{PhysRevX.10.041041,PhysRevB.105.085154,PhysRevB.102.165151,Ahn2022-pc,Resta2011, PhysRevB.62.1666, Raffaele_Resta_2002,PhysRevB.104.134312,PhysRevResearch.4.013217,2021arXiv211107782H,PhysRevB.94.134423,PhysRevLett.121.020401}.
 We prove that this quantum metric agrees with the usual definition of the quantum metric in noninteracting systems and is positive semi-definite even in an interacting system, as necessary for a metric. Furthermore, we show that the GQM is related to the Chern number at $T=0$. We then use this definition to calculate the quantum metric of an interacting system going through a topological phase transition and show that the GQM is sensitive to topological phase transitions. While previous works characterized quantum phase transitions by the quantum metric calculated from the quantum states\cite{PhysRevE.74.031123,PhysRevLett.99.100603,PhysRevB.104.045103,10.21468/SciPostPhys.12.1.018}, we utilize the Green's function to detect the transition.

The rest of the paper is organized as follows: In  section~\ref{sec:definition}, we define the generalized quantum metric in interacting systems and demonstrate its correspondence to the usual quantum metric in noninteracting systems. We furthermore prove that 
the GQM  
is positive semi-definite and relates to the Chern number at $T=0$. In section~\ref{sec:numerics}, we use this definition to study the topological phase transition in interacting systems. Finally, we summarize and conclude this paper in section~\ref{sec:conclusion}.

\section{Quantum metric in correlated systems}\label{sec:definition}
\subsection{Definition}
In previous studies~\cite{Resta2011, PhysRevB.62.1666, Raffaele_Resta_2002}, the following relationship between the quantum metric and the optical conductivity at $T=0$ has been pointed out in noninteracting systems:
\begin{eqnarray}
\mathrm{Re}\int_0^{\infty} d\omega_1 \;\frac{\sigma_{\alpha\beta}(\omega_1) + \sigma_{\beta{\alpha}}(\omega_1)}{2\pi\omega_1}&=&\frac{1}{(2\pi)^d}\int d\bk\; g_{\alpha\beta}(\bk),\nonumber\\
\label{SWM_k}
\end{eqnarray}
where $\sigma_{\alpha\beta}(\omega)$ is the optical conductivity and $d$ is the dimension of the system. Although the original work~\cite{PhysRevB.62.1666} deals with three-dimensional systems, it is easily seen that the relation holds for any dimension. Throughout this paper, we set the Planck constant and the lattice constant to unity, $\hbar=a=1$. We also set the electron charge and the Boltzmann constant
to unity,  $e = k_B = 1$. Eq.~($\ref{SWM_k}$) holds for noninteracting systems, where $g_{\alpha\beta}(\bk)$ is the existing definition of the quantum metric on the Brillouin Zone (BZ).
 
Taking Eq.~(\ref{SWM_k}) as a starting point,
 we propose a generalization of $g_{\alpha\beta}$ on the BZ. In this paper, the generalized quantum metric is called GQM. When defining the GQM, we impose the following conditions: (i) The GQM  is identical to the existing definition in noninteracting systems. (ii) The GQM is positive semi-definite even in correlated systems. (iii) The GQM can be expressed by the single-particle spectral function. (iv) The GQM is related to the Chern number. We impose condition (i) so that we can regard the GQM as a generalization of the conventional definition. The GQM should satisfy (ii) because the GQM should serve as a metric. Condition (iii) is imposed because the GQM should be based on single-particle states as in noninteracting systems. In addition, condition (iv) is imposed because the relationship between the conventional quantum metric and the Chern number has been pointed out in noninteracting systems~\cite{Peotta2015,PhysRevB.104.045103, PhysRevB.103.L241102,Zhang_2022,10.21468/SciPostPhys.12.1.018}. Although the existing quantum metric has a rich mathematical structure\cite{https://doi.org/10.48550/arxiv.2205.15353,PhysRevB.104.045104,PhysRevB.106.165133}, we do not restrict it when defining the GQM. This paper aims to provide a step forward in generalizing the quantum metric to interacting systems. 
 
To satisfy the above conditions (i)-(iv), we define the GQM by the following equation:
\begin{eqnarray}
\label{defGQM}\tilde{g}_{\alpha\beta}(\bk)&=&\mathrm{Re}\int_0^{\infty} d\omega_1 \frac{\sigma_{\alpha\beta}(\omega_1,\bk) + \sigma_{\beta{\alpha}}(\omega_1,\bk)}{2\pi\omega_1}\\
\label{optk}\sigma_{\alpha\beta}(\omega_1,\bk)&=&-\frac{1}{\omega_1}\int \frac{d\omega}{2\pi}f(\omega)\nonumber\\
&&\times \mathrm{Tr}\Bigl[J_{\alpha\beta}G^{R-A}(\omega)\nonumber\\
&& \ \ \ \ \ \ +J_\alpha G^R(\omega+\omega_1)J_\beta G^{R-A}(\omega)\nonumber\\ 
&& \ \ \ \ \ \ +J_\alpha G^{R-A}(\omega)J_\beta G^A(\omega-\omega_1)\Bigr],
\end{eqnarray}
where
 $f(\omega)=1/(\exp(\beta \omega) + 1)$ is the Fermi distribution function, $J_{\alpha}=\frac{\partial H}{\partial k_\alpha}$, $J_{\beta}=\frac{\partial H}{\partial k_\beta}$ are the current operators. $G^R$ and $G^A$ are the retarded and advanced single-particle Green's functions, respectively. We define $G^{R-A}(\omega) = G^R(\omega) - G^A(\omega)$. We omit explicitly writing the momentum dependence of $J$ and $G^{R/A}$. 
 These single-particle Green's functions describe the single-particle properties of a system even in the presence of correlations. They can be measured by experiments, such as angle-resolved photoemission spectroscopy and tunneling experiments. 
Because single-particle Green's functions can include correlation effects via the self-energy, this equation does describe not only noninteracting systems but also correlated systems. The optical conductivity is expressed as the integral of $\sigma_{\alpha\beta}(\omega_1,\bk)$ ~\cite{PhysRevB.103.195133}: 
\begin{eqnarray}
\sigma_{\alpha\beta}(\omega_1) &=& \frac{1}{(2\pi)^d}\int d\bk \sigma_{\alpha\beta}(\omega_1,\bk)\label{opt}
\end{eqnarray}
We note, however, that vertex corrections are ignored when expressing the optical conductivity by single-particle Green's functions in Eq.~(\ref{defGQM}) and Eq.~(\ref{opt}).
We will show next that Eq.~(\ref{defGQM}) corresponds to the usual definition of the quantum metric in noninteracting systems and is positive semi-definite as necessary for a metric. Furthermore, we will point out the relationship between the GQM and Chern number at $T=0$. 
\subsection{Correspondence to noninteracting systems}
First, we will show that the GQM coincides with the conventional quantum metric in noninteracting systems. 

Because the first term in Eq.~(\ref{optk}) is generally purely imaginary, it can be ignored.
In a noninteracting system, the Hamiltonian can be diagonalized by single-particle wave functions $\vert n\rangle$ yielding $H\vert n\rangle =E_n\vert n\rangle$, and the Green's function can also be diagonalized by the same eigenvector as $G^{R/A}\ket{n} = (\omega-E_n \pm i\eta)^{-1}\ket{n}$. Because $\eta\rightarrow0$ holds
in noninteracting systems,
$G^{R-A}\ket{n} = -2\pi i\delta(\omega-E_n)\ket{n}$.
Using these eigenfunctions and equations, we can write the second term in Eq.~(\ref{optk}) as
\begin{eqnarray}
&&\Tr{J_\alpha G^R(\omega+\omega_1)J_\beta G^{R-A}(\omega)}\nonumber\\
&&=-2\pi i \sum_{n,m}\frac{ (J_\alpha)_{nm}(J_\beta)_{mn}}{\omega + \omega_1 -E_m + i\eta}\delta(\omega - E_n).\label{first}
\end{eqnarray}
Similarly,
\begin{eqnarray}
&&\Tr{J_\alpha G^{R-A}(\omega) J_\beta G^A(\omega-\omega_1)} \nonumber\\
&&=2\pi i \sum_{n,m}\frac{ (J_\alpha)_{nm}(J_\beta)_{mn}}{\omega - \omega_1 -E_n - i\eta}\delta(\omega - E_m).\label{second}
\end{eqnarray}
Inserting Eq.~(\ref{first}) and (\ref{second}) into Eq.~(\ref{optk}),
 we obtain
\begin{align}
&\ReB{\sigma_{\alpha\beta}(\omega_1,\bk)}\nonumber\\
&=  \frac{1}{\omega_1}\sum_{nm}\Bigl(f(E_n)-f(E_m)\Bigr)\mathrm{Re}\Bigl[\frac{i(J_\alpha)_{nm}(J_\beta)_{mn}}{\omega_1-(E_m-E_n) + i\eta}\Bigr].
\end{align}
Symmetrizing this formula, we find
\begin{align}
&\ReB{\frac{\sigma_{\alpha\beta}(\omega_1,\bk) + \sigma_{\beta \alpha}(\omega_1,\bk)}{2}}\nonumber\\
&= \frac{1}{2\omega_1}\sum_{n\neq m}\Bigl(f(E_n)-f(E_m)\Bigr)\nonumber\\
& \ \ \ \ \times\mathrm{Re}\Bigl[\frac{i((J_\alpha)_{nm}(J_\beta)_{mn}+(J_\beta )_{nm}(J_\alpha)_{mn})}{\omega_1-(E_m-E_n) + i\eta}\Bigr]
\end{align}
By using
$(J_\alpha)_{nm}(J_\beta)_{mn}+(J_\beta )_{nm}(J_\alpha)_{mn} = 2\mathrm{Re}[(J_\alpha)_{nm}(J_\beta)_{mn}]$, 

\begin{align}
&\ReB{\frac{\sigma_{\alpha\beta}(\omega_1,\bk) + \sigma_{\beta \alpha}(\omega_1,\bk)}{2}}\nonumber\\
&=\frac{\pi }{\omega_1}\sum_{n\neq m} \delta(\omega_1-(E_m-E_n))\Bigl(f(E_n)-f(E_m)\Bigr)\nonumber\\
& \ \ \ \ \ \ \ \ \ \ \ \ \label{fnfm}\times\ReB{(J_\alpha)_{nm}(J_\beta)_{mn}}
\end{align}
For $T=0$, this can be written as
\begin{eqnarray}
\frac{\pi}{2\omega_1}\displaystyle \sum_{n:E_n<0}\displaystyle \sum_{m:E_m>0}((J_\alpha)_{nm}(J_\beta)_{mn}+(J_\beta )_{nm}(J_\alpha)_{mn})\nonumber \\
\times [\delta(\omega_1-(E_m-E_n))-\delta(\omega_1-(E_n-E_m))].
\end{eqnarray}
Finally, we can write the GQM as 
\begin{eqnarray}
\tilde{g}_{\alpha\beta}(\bk) &=& \mathrm{Re} \;\int_{0}^{\infty}d\omega_1\frac{\sigma_{\alpha\beta}(\omega_1,\bk) + \sigma_{\beta{\alpha}}(\omega_1,\bk)}{2\pi\omega_1}\\
&=& \frac{1}{2}\displaystyle \sum_{n:E_n<0}\displaystyle \sum_{m:E_m>0} \Bigl[\frac{\langle n|\partial_\alpha H |m\rangle\langle m|\partial_\beta H |n\rangle}{(E_n-E_m)^2}\nonumber\\
&& \ \ \ \ \ \  + (\alpha \leftrightarrow \beta)\Bigr] \nonumber\\
&=& \mathrm{Re} \displaystyle \sum_{E_n<0}\displaystyle \sum_{E_m>0} \frac{\langle n|\partial_\alpha H |m\rangle\langle m|\partial_\beta H |n\rangle}{(E_n-E_m)^2},\label{existing}
\end{eqnarray}
which agrees with the definition of the quantum metric in noninteracting systems~\cite{PhysRevB.56.12847}.

\subsection{Proof of positive semi-definiteness}
As a metric, the GQM must be positive semi-definite. This must hold in interacting systems as well.
The strategy employed here is to use the eigenstates of $G^{R-A}(\omega)$ (not $G^R$ or only $G^A$). This can be done by symmetrizing with respect to $\alpha$ and $\beta$. For example, after symmetrizing Eq.~(\ref{second}),
\begin{align}
&\ReB{\Tr{J_\alpha G^R(\omega+\omega_1)J_\beta G^{R-A}(\omega) +(\alpha \leftrightarrow \beta)}}\nonumber\\ 
&= \ReBL{\TrL{J_\alpha G^R(\omega+\omega_1)J_\beta G^{R-A}(\omega)}}\nonumber\\
& \ \ \ \ \ \ \ \ \ \ \ \ \ \ + \{J_\beta G^R(\omega+\omega_1)J_\alpha G^{R-A}(\omega)\}^{\dagger}\Bigr]\Bigr]\\
&= \ReB{\Tr{J_\alpha G^{R-A}(\omega+\omega_1)J_\beta G^{R-A}(\omega)}}\label{GRGA}
\end{align}
Since $G^{A} = (G^{R})^{\dagger}$, $i(G^{R} - G^{A})$ is a Hermitian matrix. Thus, $i(G^{R} - G^{A})$ is diagonalizable by a unitary matrix, and the eigenvalues are real. 

We denote the eigenvectors as $\ket{n(\omega)}$, and the corresponding eigenvalues as $A_n(\omega) (\in\mathbb{R})$. Note that $A_n(\omega) \geq 0$ because $G^{R-A}(\omega)$ corresponds to the spectral function. Then, Eq.~(\ref{GRGA}) is equal to 
\begin{eqnarray}
&& \displaystyle \sum_{n,m}-A_n(\omega)A_m(\omega+\omega_1)\times \Bigl(\langle n(\omega)|J_\alpha|m(\omega + \omega_1)\rangle\nonumber\\
&&\times \langle m(\omega+\omega_1)|J_\beta|n(\omega)\rangle 
+ (\alpha\leftrightarrow\beta) \Bigr)/2.
\end{eqnarray}
For the third term in Eq.~(\ref{optk}), we utilize the same method, substituting

\begin{align}
 \omega-\omega_1 \rightarrow \omega,\ \ \  f(\omega)\rightarrow f(\omega + \omega_1)
\end{align}

 Finally, we obtain all terms which contribute to the GQM. The argument for the frequency integration can be written as
 \begin{eqnarray}
&&\displaystyle \sum_{n,m}A_n(\omega)A_m(\omega+\omega_1)\nonumber\\
&& \times \Bigl(\langle n(\omega)|J_\alpha|m(\omega + \omega_1)\rangle\langle m(\omega+\omega_1)|J_\beta|n(\omega)\rangle \nonumber\\
&&+ (\alpha\leftrightarrow\beta) \Bigr)\times \frac{f(\omega)-f(\omega+\omega_1)}{\omega_1^2}\times \frac{1}{8\pi^2}.
\end{eqnarray}
We define
\begin{align}
h(n,m,\omega,\omega_1) = A_n(\omega)A_m(\omega+\omega_1)\frac{f(\omega)-f(\omega+\omega_1)}{8\pi^2\omega_1^2}
\end{align}
It is important to notice that $h(n,m,\omega,\omega_1)$ is non-negative because of the positive semi-definiteness of the spectral function and the monotonicity of the Fermi distribution function.
Then, for arbitrary real numbers $c_\alpha$,
\begin{align}
\displaystyle &\sum_{\alpha,\beta}c_\alpha \tilde{g}_{\alpha\beta}c_{\beta}\nonumber\\
&= \displaystyle 2\sum_{n,m,\alpha,\beta}\int_{0}^{\infty}d\omega_1\int_{-\infty}^{\infty}d\omega \;h(n,m,\omega,\omega_1)\;\nonumber\\
& \ \ \ \times \langle n(\omega)|c_\alpha J_\alpha|m(\omega + \omega_1)\rangle \langle m(\omega+\omega_1)|c_\beta J_\beta|n(\omega)\rangle.\label{PSD}
\end{align}
Denoting
\begin{align}
|\phi_n(\omega)\rangle = \displaystyle \sum_{\alpha} c_\alpha J_\alpha|n(\omega)\rangle ,
\end{align}
we complete the proof of the positive semi-definiteness;
\begin{eqnarray}
\displaystyle \sum_{\alpha,\beta}c_\alpha \tilde{g}_{\alpha\beta}c_{\beta}&=&\displaystyle 2\sum_{n,m} \int_{0}^{\infty}d\omega_1\int_{-\infty}^{\infty}d\omega h(n,m,\omega,\omega_1)\nonumber\\
&& \ \times \langle \phi_n(\omega)| m(\omega+\omega_1)\rangle\langle m(\omega+\omega_1)|\phi_n(\omega)\rangle \nonumber\\
&\geq& 0
\end{eqnarray}
Thus, the GQM serves as a metric even if correlation effects are present.
As we can see from this proof, the GQM is expressed as
\begin{eqnarray}
    \tilde{g}_{\alpha\beta}&=&\displaystyle \sum_{n,m}\int_{0}^{\infty}d\omega_1\int_{-\infty}^{\infty}d\omega A_n(\omega)A_m(\omega+\omega_1)\nonumber\\
&& \times \Bigl(\langle n(\omega)|J_\alpha|m(\omega + \omega_1)\rangle\langle m(\omega+\omega_1)|J_\beta|n(\omega)\rangle \nonumber\\
&&+ (\alpha\leftrightarrow\beta) \Bigr)\times \frac{f(\omega)-f(\omega+\omega_1)}{\omega_1^2}\times \frac{1}{8\pi^2}\label{GQM_A}.
\end{eqnarray}
The GQM is described by
the single-particle spectral function and its eigenvectors and does not include the Bloch wave function. This point is the major difference between the GQM and the dressed quantum metric in Ref.~\cite{10.21468/SciPostPhysCore.5.3.040}. We utilize the eigenvectors of the single-particle spectral function. They coincide with the Bloch wave functions in noninteracting systems but are generally different in interacting systems. The GQM can capture the feature of the single-particle states even in correlated systems.

\subsection{Relation to the Chern number}

In two-dimensional systems, the following equation and inequality for the quantum metric on the twist-angle space are satisfied at $T=0$:
\begin{eqnarray}
    &&\mathrm{Re} \int^\infty_0 d\omega_1 \frac{\sigma_{xx}(\omega_1) +\sigma_{yy}(\omega_1)}{\pi\omega_1}\nonumber\\
    &&= \frac{1}{(2\pi)^2}\int d\bm{\kappa} \mathrm{Tr}g_{\mathrm{TBC}}(\bm{\kappa})\label{optk_twist_utilize}\\
    &&\geq\frac{1}{(2\pi)^2}\int d\bm{\kappa}|\Omega_{\mathrm{TBC}}(\bm{\kappa})|\label{trqm_Berry}\\
    &&\geq\frac{1}{2\pi}|Ch|,
\end{eqnarray}
where $g_{\mathrm{TBC}}(\kappa)$ and $\Omega_{\mathrm{TBC}}(\bm{\kappa})$ are the quantum metric and the Berry Curvature introduced by imposing twisted boundary conditions, respectively~\cite{Resta2011, PhysRevB.62.1666, Raffaele_Resta_2002}. Eq.~(\ref{optk_twist_utilize}) is proportional to the localization tensor, which measures the degree of the localization of the wave function~\cite{PhysRevB.62.1666}. $Ch$ is the Chern number determined by $\Omega_{\mathrm{TBC}}(\bm{\kappa})$. This Chern number is the so-called Niu-Thouless-Wu 
 invariant, which contributes to the quantized Hall conductivity in interacting systems~\cite{PhysRevB.31.3372}. Eq.~(\ref{optk_twist_utilize}) 
 has been pointed out in Ref.~\cite{PhysRevB.62.1666} assuming  the ground state is non-degenerate and gapped for any twist angle $\bm{\kappa}$. When deriving the inequality Eq.~(\ref{trqm_Berry}), the general inequality $\mathrm{Tr}g_{\mathrm{TBC}}(\bm{\kappa})\geq |\Omega_{\mathrm{TBC}}(\bm{\kappa})|$ is used \cite{Peotta2015}. 
Thus, the GQM is also expected to be large in Chern insulators at $T=0$ due to the following inequality:
\begin{eqnarray}
    \frac{1}{(2\pi)^2}\int d\bk \mathrm{Tr}\tilde{g}&\approx& \mathrm{Re} \int^\infty_0 d\omega_1 \frac{\sigma_{xx}(\omega_1) +\sigma_{yy}(\omega_1)}{\pi\omega_1} \label{vertex_ignore}\\
    &\geq&\frac{1}{2\pi}|Ch|\label{trivial}
\end{eqnarray}
We note that this inequality is not rigorous because vertex corrections are ignored in Eq.~(\ref{vertex_ignore}). In the next section, we numerically verify the relationship between the GQM and the Chern number and check the above inequality.

\section{Analysis of correlation effects on the GQM}\label{sec:numerics}
In this section, we numerically confirm the properties of the GQM pointed out in the previous section and investigate the relationship between the GQM and the Chern number. For this purpose, we calculate the GQM for the following Hamiltonian:

\begin{eqnarray}
H &=& H_{\mathrm{QWZ}} + H_{\mathrm{int}}\\
H_{\mathrm{QWZ}} &=& \displaystyle \sum_{\bk} (c^{\dagger}_{\bk,A}\; c^{\dagger}_{\bk,B})\;H(\bk)\;(c_{\bk,A}\; c_{\bk,B})^{T}\\
H(\bk) &=& \sin k_x\sigma_x + \sin k_y\sigma_y \nonumber\\
&& \ \ \ + (M+\cos k_x +\cos k_y)\sigma_z\\
H_\mathrm{int} &=& U\displaystyle \sum_i n_{i,A}n_{i,B}, 
\end{eqnarray}
where $A$ and $B$ denote pseudo spin indices, $i$ labels the sites on the lattice, and $\sigma_i$ are the Pauli matrices. $M$ plays the role of a staggered potential~\cite{asboth2016short}(or magnetic field). $H_\mathrm{QWZ}$ is known as the QWZ model, which is a toy model of a Chern insulator on a square lattice introduced by  X.-L. Qi, Y.-S. Wu, and S.-C. Zhang~\cite{PhysRevB.74.085308,asboth2016short}.  
The Chern number of this model is $Ch=1$ when $0 < M < 2$ and $Ch=-1$ when $-2 < M < 0$. This model is topologically trivial if $\vert M\vert >2$. $M=\pm 2$ and $M=0$ are quantum critical points, for which the gap at the Fermi energy closes ~\cite{asboth2016short}.
In our calculations, we set the chemical potential as $\mu=-U/2$, corresponding to a half-filled system.

\subsection{$U=0$ cases}

Before investigating the $U$-dependence of the GQM, we confirm the correspondence to the existing definition of the quantum metric in the noninteracting system in Eq.~(\ref{existing}). 
 When evaluating the GQM, all calculations are performed at $T=0$. 
 In our numerical calculations, we use $\eta = 0.02$ in the noninteracting Green's function, $G^{R/A} = (\omega-H(\bk) \pm i\eta)^{-1}$.
$\eta$ can be regarded as an inverse lifetime of the quasiparticles due to scattering in the model. 
Thus, the spectral function at the Fermi energy becomes finite even in the insulating states due to the finiteness of $\eta$. This results in a finite DC conductivity, leading to a divergence of the GQM. To avoid such an artificial divergence, the frequency integral in Eq.~(\ref{defGQM}) needs to be cut at a small frequency of the order of $\eta$. In this paper, we have used $\omega_0 = 0.01$.

\begin{figure}[t]
\centering
\includegraphics[width=\linewidth]{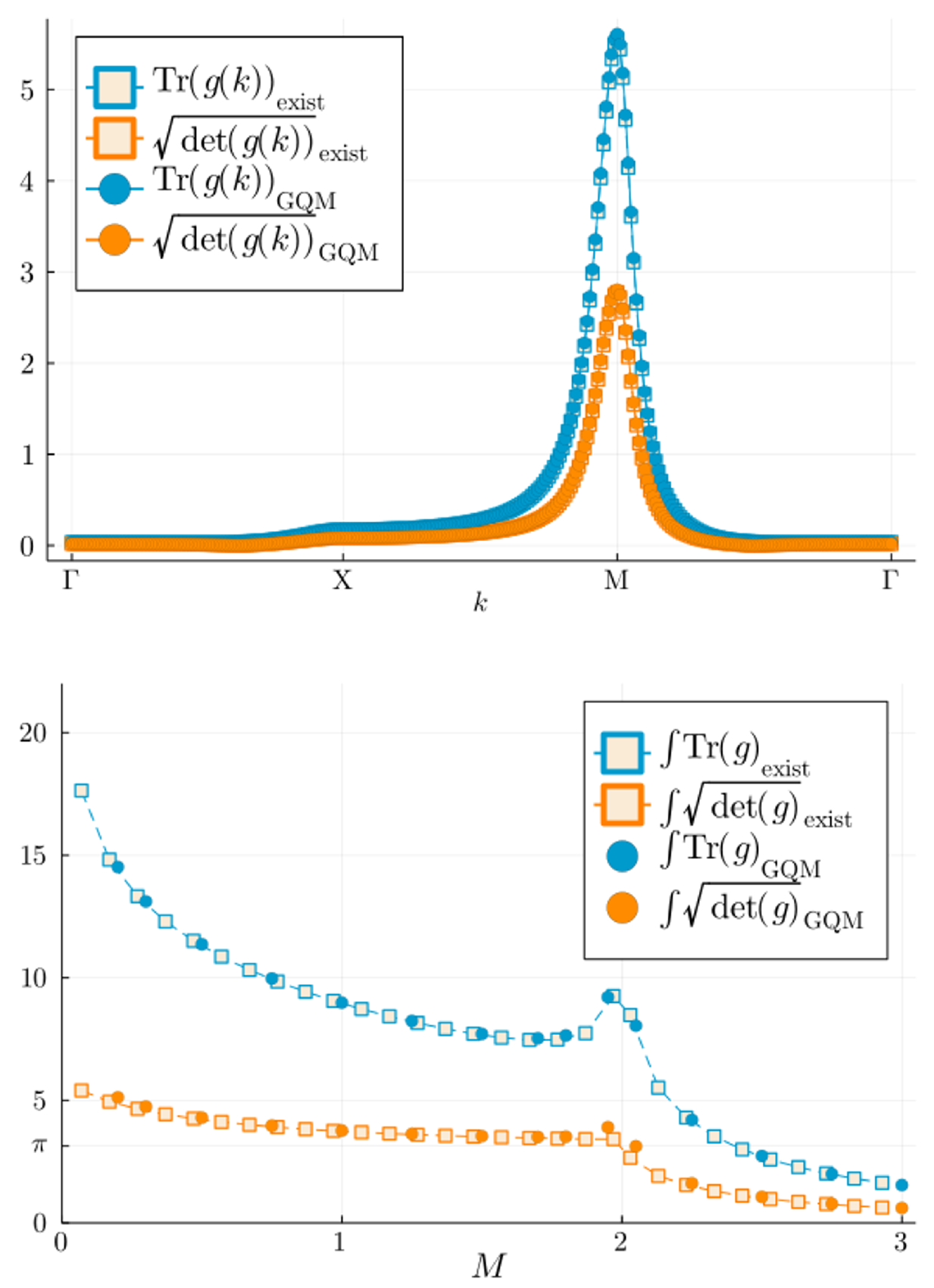}
\caption{top: The momentum-dependence of the quantum metric calculated by the existing formula (squares) compared to the GQM (Eq.~(\ref{defGQM})) (circles) at $M=1.7$ and $U=0$. bottom: $M$-dependence of the integral of the quantum metric calculated by the existing formula (squares) compared to the GQM (Eq.~(\ref{defGQM})) (circles) at $U=0$. We note that we do not explicitly show $M=2$ as the existing definition diverges at this point.}
\label{fig:qmcompare}
\end{figure}

In Fig.~\ref{fig:qmcompare}(top), we compare the momentum-dependence of $\mathrm{Tr}(g(\bk))$ and $\sqrt{\mathrm{det}(g(\bk))}$ calculated by the existing definition and the GQM. We see that they coincide very well, but the GQM is a little larger, particularly near the $\mathrm{M}$ point, where the gap is smallest. 
This enhancement can be traced back to $\eta$ in the noninteracting Green's function. (A more detailed analysis of the $\eta$ dependence is given in Appendix \ref{appendix_dissipation}.)

This broadening also results in a change of the behavior of $\detg$, which is called quantum volume in Ref.~\cite{PhysRevB.104.045103}, in the noninteracting system, shown in Fig.~\ref{fig:qmcompare}(bottom). Near the topological phase transition, $M=2$, a small peak is observed for $\int \sqrt{\mathrm{det}(g)}_\mathrm{GQM}dk_xdk_y$ while the conventional definition does not show such a peak.

 Furthermore, peaks in $\int \mathrm{Tr}(g)dk_xdk_y$  of the GQM are observed at the critical points ($M=0,\;2$). The reason for these peaks is a finite longitudinal optical conductivity at small $\omega$. 
 We note that we omitted the point at $M=2$, where the existing definition of $\trg$ diverges.
Overall, we see that the existing definition of the quantum metric and the GQM agree very well for small $\eta$.
 
In addition, Fig.~(\ref{fig:qmcompare})(bottom) shows that the inequality $\detg \geq \pi |Ch| = \pi$ pointed out 
in Ref.~\cite{PhysRevB.104.045103} is satisfied even for the GQM in the topological phase. 
This inequality can be used to distinguish the topologically trivial phase from the topologically nontrivial phase in the QWZ model,\cite{PhysRevB.104.045103}
and we will later show this detection is possible even in the interacting QWZ model by using the GQM.
Furthermore, we see that $\trg$ and $\detg$ tend to be larger in the topological phase than in the trivial phase. (The same behavior for $\detg$ has been confirmed numerically in various models of Chern insulators~\cite{PhysRevB.104.045103}.) 

\subsection{$U\neq0$ cases}
Next, we focus on the interacting system and show the results for 
$U>0$.
We analyze the effect of interaction by using the dynamical mean-field theory (DMFT). DMFT self-consistently maps  a lattice model to an impurity model~\cite{RevModPhys.68.13}. DMFT neglects the momentum dependence of the self-energy but incorporates the energy dependence of it ~\cite{RevModPhys.68.13}. It has been frequently utilized to analyze the topology of correlated electron systems~\cite{PhysRevB.87.085134,PhysRevB.85.125113,PhysRevB.87.235104,PhysRevB.85.165138,PhysRevB.103.165130,PhysRevB.85.235135,PhysRevX.9.041055,PhysRevB.93.235159,PhysRevB.98.075104}. 
To solve the impurity model, we utilize the numerical renormalization group (NRG)~\cite{RevModPhys.80.395,PhysRevB.74.245114}.

Besides using the GQM, we use Wilson Loops to analyze the topology of the system\cite{asboth2016short,PhysRevB.99.045140,PhysRevB.100.195135} focusing on the effective Hamiltonian~\cite{PhysRevX.2.031008,article}
\begin{eqnarray}
H_{\mathrm{eff}}(\bk) = H(\bk) + \Sigma^R(\omega = 0), 
\end{eqnarray}
where $\Sigma^R(\omega)$ is the self-energy of the retarded Green's function calculated by DMFT. Because $\omega=0$ lies in the gap in our setup, this effective Hamiltonian can detect the topology of the system. 
In the numerical calculations, we employ the discretized formula of the Wilson Loop, 
\begin{eqnarray}
W(k_x) =\langle u(k_x,2\pi)| \displaystyle \prod_{n=1}^{N}P(k_x,2\pi n/N) |u(k_x, 2\pi)\rangle,
\end{eqnarray}
where $|u(k_x,k_y)\rangle$ is the occupied eigenstate of $H_{\mathrm{eff}}(\bk)$ and $P(\bk)=|u(k_x,k_y)\rangle\langle u(k_x,k_y)|$  is the projection operator. The product is arranged so that $k_y$ increases when going from the left to the right. Furthermore,  $N$ must be large enough in a numerical calculation. A finite winding number is a signal of a nontrivial topology of the filled bands~\cite{asboth2016short,PhysRevB.99.045140,PhysRevB.100.195135}. If the winding number is zero, the topology of the bands can be judged to be trivial. 
Because the imaginary part of the self-energy at the Fermi energy is almost zero (a typical value is $10^{-4}$) in the parameter region analyzed in our numerical calculations, we can neglect the imaginary part. Thus, the effective Hamiltonian is Hermitian, and we can utilize the eigenvectors corresponding to the smallest eigenvalues when calculating Wilson loops.

It is expected that the system will undergo a topological phase transition as $U$ increases. This phase transition can be understood as follows: The self-energy affects the band structure around the Fermi energy by changing the staggered potential to 
$M_{\mathrm{eff}}=M+(\mathrm{Re}\Sigma^R(\omega = 0)_{AA}-\mathrm{Re}\Sigma^R(\omega = 0)_{BB})/2$, where  $\mathrm{Re}\Sigma^R(\omega = 0)_{AA}$ and $\mathrm{Re}\Sigma^R(\omega = 0)_{BB}$ are the real parts of the diagonal elements of the self-energies. Thus, we expect topological phase transitions in the interacting system when $M_{\mathrm{eff}}=\{-2,0,2\}$.

\begin{figure}[t]
\centering
\includegraphics[width=\linewidth]{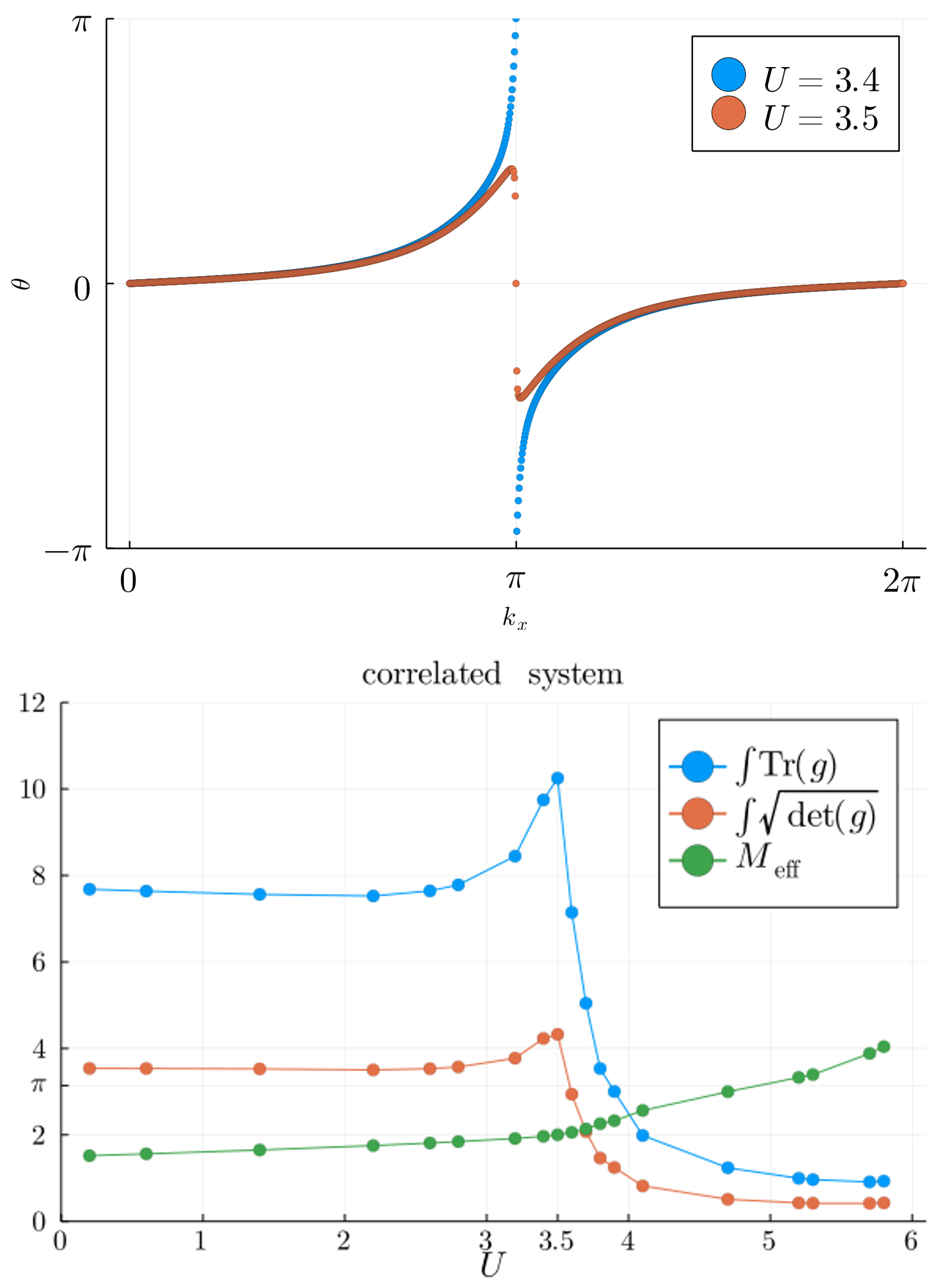}
\caption{top: The $k_x$-dependence of the phase of $W(k_x)$ 
at $M=1.5$ for two different interaction strengths. The Wilson loop is winding for $U = 3.4$ (blue dots). The Wilson loop is not winding for $U = 3.5$ (orange dots).  bottom: Interaction dependence of  $\trg$(blue line) and $\detg$(red line) calculated by Eq.~(\ref{defGQM}). At the transition point, the effective staggered potential (green line)  $M_{\mathrm{eff}} = M+(\mathrm{Re}\Sigma^R(\omega = 0)_{AA}-\mathrm{Re}\Sigma^R(\omega = 0)_{BB})/2 = 2$, which is the same value as in the noninteracting system. The parameters are $M=1.5,\mu = -U/2,\eta=0.02$.}
\label{fig:Wilson_M=1.5}
\end{figure}

In fact, for $U=3.5$, the system undergoes a topological phase transition from the topological phase to the trivial phase as $U$ increases. This is confirmed by the Wilson Loop, shown in Fig.~\ref{fig:Wilson_M=1.5}(top). The winding number of the Wilson Loop is one when $U < U_c \approx 3.5$, and it is zero when $U > U_c$. 

In Fig.~\ref{fig:Wilson_M=1.5}(bottom), we show the corresponding GQM.
We see a small peak in $\int \mathrm{Tr}(g)dk_xdk_y$ and $\int \sqrt{\mathrm{det}(g)}dk_xdk_y$ at $U=3.5$ indicating the topological phase transition.
This behavior is similar to that in the noninteracting system. A small peak in
$\trg$ also appears in a correlated system at the topological phase transition. The reason for this small peak is the finite DC conductivity due to the gap closing.
Remarkably, the inequality $\detg \geq \pi$ still holds in the topological phase of the correlated system. Due to the general inequality for the positive semi-definite tensor $\mathrm{Tr}(g(\bk))\geq 2\sqrt{\mathrm{det}(g(\bk))}$, we see that $\trg\geq 2\pi$ holds for the topological phase. This corresponds to the inequality  Eq.~(\ref{trivial}). Although vertex corrections are ignored when deriving Eq.~(\ref{trivial}), the results show that the inequality holds even in the interacting systems in this model. In addition, $\int \mathrm{Tr}(g)dk_xdk_y$ and $\int \sqrt{\mathrm{det}(g)}dk_xdk_y$ are larger in the topological phase compared to the trivial phase. 
Thus, similar to the noninteracting system, $\detg\geq \pi$ can be used as a probe to judge the topology of the system.

We believe that the small peak in $\detg$ at $U=3.5$ is caused by the small $\eta$.
We note that the behavior of $\detg$ in the correlated system near the transition point is similar to that in the noninteracting system.  
The effects of $\eta$ on the GQM are also investigated in Appendix \ref{appendix_dissipation}.

\begin{figure}[t]
\centering
\includegraphics[width=\linewidth]{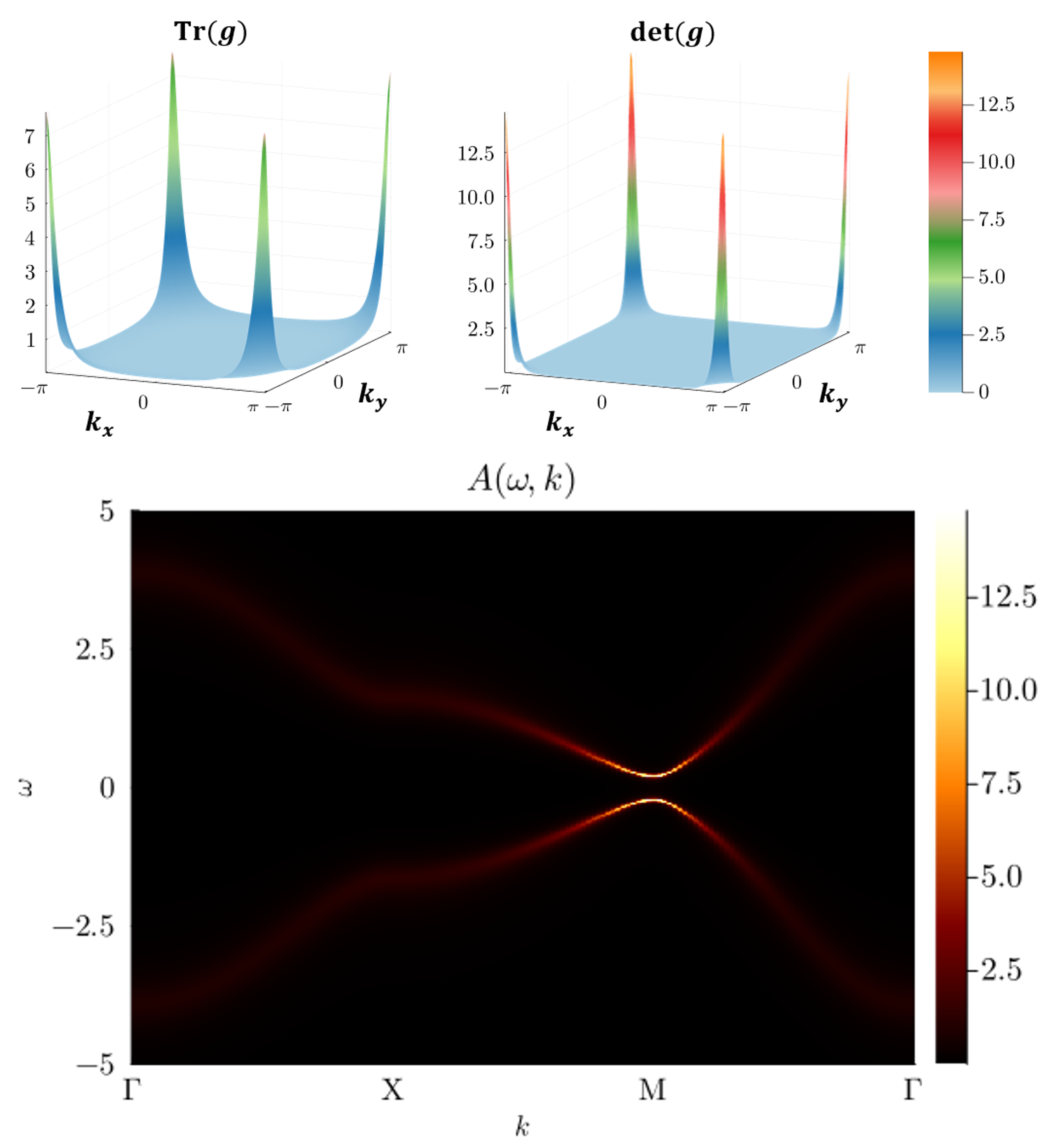}
\caption{top: Momentum-dependence of $\mathrm{Tr}g(\bk)$ and $\mathrm{det}g(\bk)$ demonstrating the positive semi-definiteness of the GQM. bottom: Spectral function $A(\omega,\bk)=-\frac{1}{\pi}\text{Tr}\;\text{Im}G^R(\omega,\bk)$. The gap becomes small near the $\mathrm{M}$ point. In both figures, the parameters are $M=1.5,U=2.2,\mu = -U/2,\eta=0.02$.}
\label{fig:trmapM=1.5}
\end{figure}

Next, we demonstrate that the GQM is positive semi-definite in the correlated system.
To demonstrate this, we show the momentum dependence of $\mathrm{Tr}g(\bk)$ and $\mathrm{det}g(\bk)$ in Fig.~\ref{fig:trmapM=1.5}(top). 
We note that in a two-dimensional system, the positive semi-definiteness is equivalent to $\mathrm{Tr}g(\bk) \geq 0$ and $\mathrm{det}g(\bk)\geq 0$ for any $\bk$. 

In Fig.~\ref{fig:trmapM=1.5}(top), we see that the GQM becomes large near the $M$ point ($k_x = k_y = \pi$). This is understood from the behavior of the spectral function shown in Fig.~\ref{fig:trmapM=1.5}(bottom). The energy gap
is small in this region. 
Thus, $\sigma_{\alpha\beta}(\omega_1,\bk)$ becomes finite at small $\omega$ and the GQM becomes large due to the factor $1/\omega_1$ 
in Eq.~(\ref{defGQM}).

\begin{figure}[t]
\centering
\includegraphics[width=\linewidth]{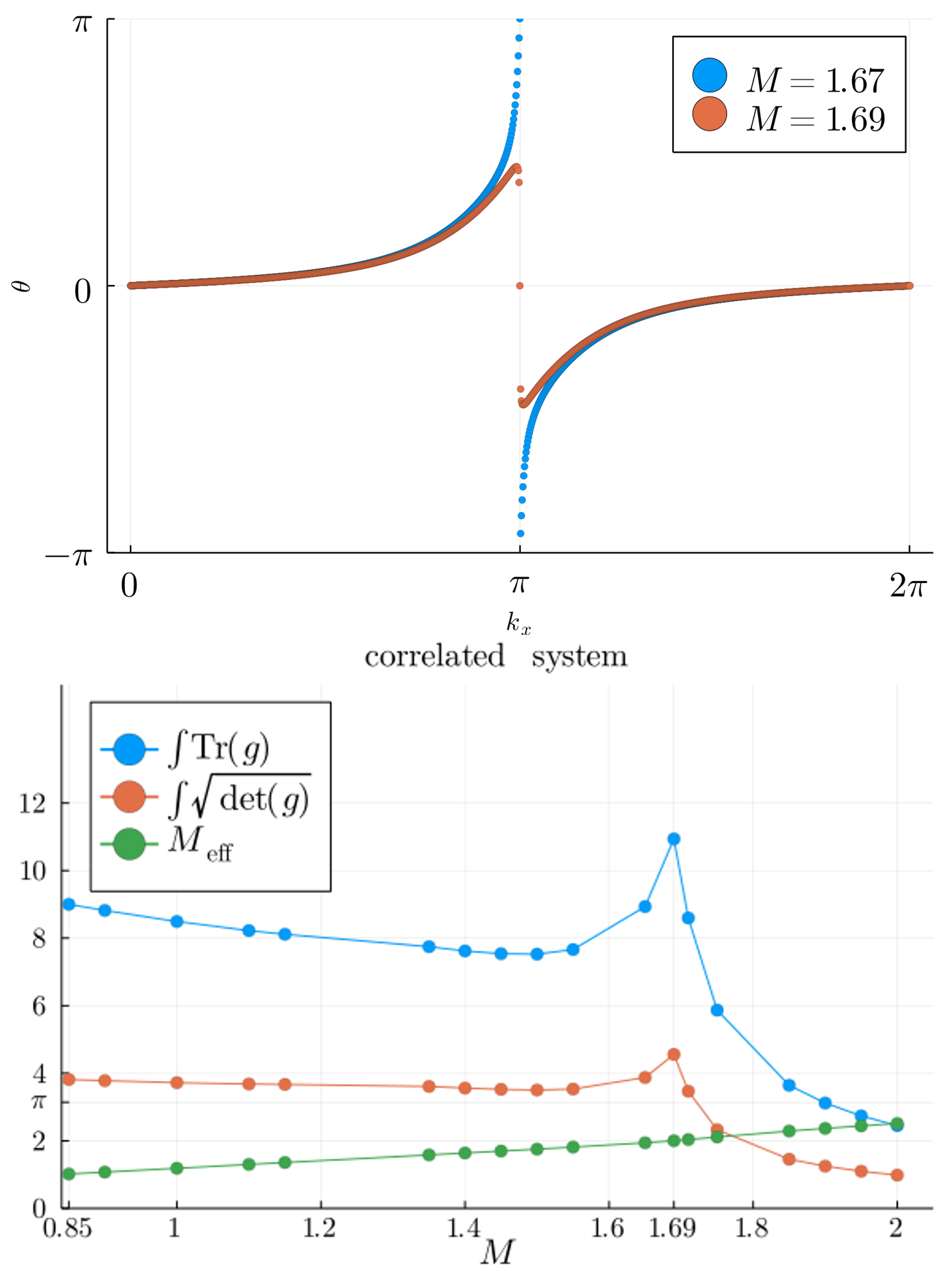}
\caption{ top: The $k_x$-dependence of the phase of $W(k_x)$
at $U = 2.2$. It is winding when $M=1.67$(blue dots) and is not winding 
when $M = 1.69$(orange dots). bottom: The $M$-dependence of the $\trg$(blue line) and $\detg$(red line) calculated by Eq.~(\ref{defGQM}). We also include $M_{\mathrm{eff}}$ as a green line, indicating that the phase transition occurs at $M_{\mathrm{eff}}=2$. The parameters are $U=2.2,\mu = -U/2,\eta=0.02$.}
\label{fig:WilsonU=2.2}
\end{figure}
Finally, we repeat our analysis for $U=2.2$ varying $M$.
The topology of the system changes as $M$ is increased. From the Wilson loop, shown in Fig.~\ref{fig:WilsonU=2.2}(top), we see that a topological phase transition from the topological phase to the trivial phase occurs at $M\approx 1.69$. Because of the presence of interactions, the value at which the topological phase transition occurs is reduced in the correlated system.

The corresponding $M$ dependence of the GQM is shown in Fig.~\ref{fig:WilsonU=2.2}(bottom). We see that the GQM shows a small peak again exactly at the topological phase transition. Furthermore, from the comparison between the values of the GQM on both sides of the topological phase transition, we infer that the phase for $M<1.69$ is topological nontrivial, in agreement with the Wilson loop analysis. We see that the inequalities $\detg \geq \pi$ and $\trg \geq 2\pi $ hold again in the whole parameter region of the topological phase.

\section{Conclusions}\label{sec:conclusion}

In summary, we have defined a generalized quantum metric(GQM) on the BZ using the single-particle Green's function in interacting systems. The GQM is based on the optical conductivity of the system. The major difference from the previous work is that the GQM is written by the eigenvectors of the spectral function, not the Bloch wave functions.
Correlation effects are included via the self-energy in the single-particle Green's function. We have shown analytically that the GQM agrees with the existing definition in a noninteracting system at zero temperature and is positive semi-definite as necessary for a metric. Furthermore, we have pointed out the relationship between the GQM and the Chern number ignoring vertex corrections.

To numerically verify our results, we have analyzed the QWZ model, a toy model of a Chern insulator, with additional repulsive interaction.
We have used the DMFT and NRG to obtain the self-energy of the system and calculate the GQM. To further diagnose the topology of the system, we have computed the Wilson Loop of the effective Hamiltonian at the Fermi energy. 

With the help of the Wilson Loop analysis, we have found that the GQM tends to be larger in the topologically nontrivial phase than in the trivial phase. In addition, the inequality between the quantum volume and the Chern number still holds in the correlated system. Thus, the GQM inherits the properties of the conventional quantum metric even in interacting systems.
We expect that the GQM affects the physical quantity as in free systems, and this point is left as future work.

\section*{Acknowledgements}
T. K deeply appreciates Ken Mochizuki, Tomoki Ozawa, Taisei Kitamura, and Shuntaro Sumita for their fruitful discussions. Y. M. is supported by RIKEN Special Postdoctoral Researcher Program and was partially supported by the WISE program. R.P. is supported by JSPS, KAKENHI Grant No. JP18K03511. 
The computation in this work has been done using the facilities of the Supercomputer Center, the Institute for Solid State Physics, and the University of Tokyo.

\appendix
 
\section{Effect of dissipation on the GQM}
\label{appendix_dissipation}
\begin{figure}[H]
\centering
\includegraphics[width=\linewidth]{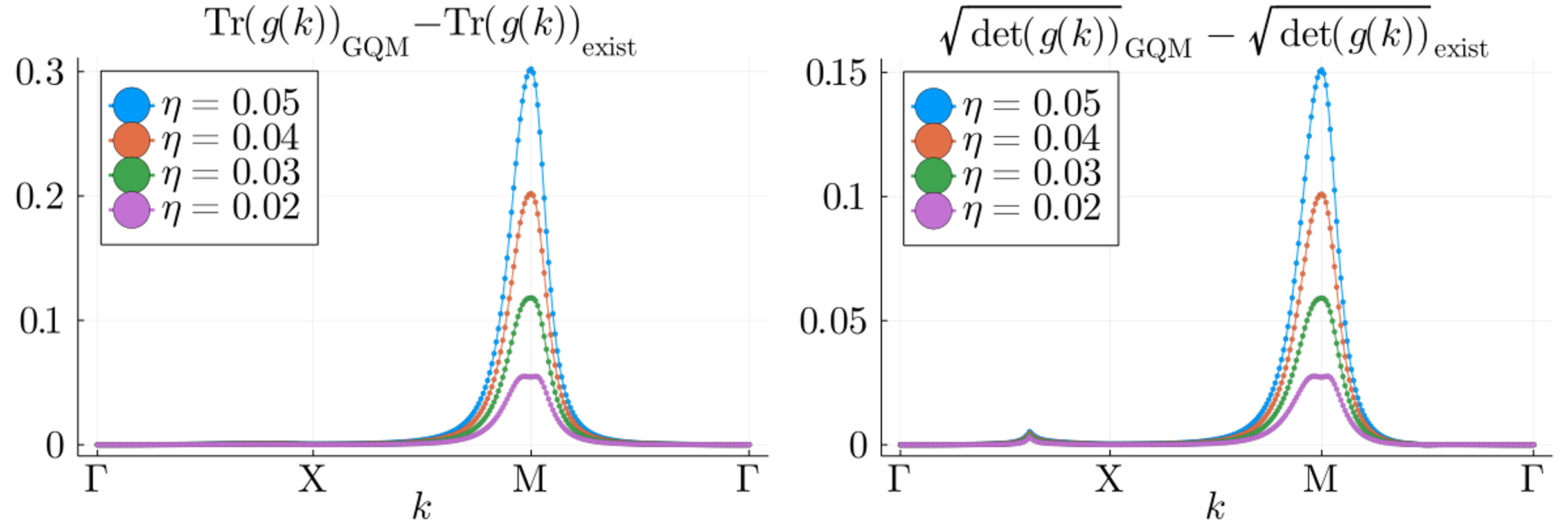}
\caption{Momentum resolved GQM in the noninteracting system ($M=1.7$ in the Hamiltonian) for different $\eta$. In these figures, the difference between the GQM and the existing definition is shown.
The left (right) panel shows $\Tr{g(k)}$ ($\sqrt{\text{det}(g(k)}$).}
\label{fig:Fig5}
\end{figure}

Although $\eta$ should be infinitesimally small, it is finite in numerical calculations. In a noninteracting system, if $\eta$ is infinitesimally small, the integrand in Eq.~\ref{defGQM} becomes finite only when $\omega=E_n$ and $\omega+\omega_1=E_m$ due to the factor $A_n(\omega)A_m(\omega+\omega_1)$ in Eq.~\ref{GQM_A}. It follows that the real part of the symmetrized optical conductivity becomes finite only when the input frequency $\omega_1$ is equal to $E_m-E_n$. However, when $\eta$ is finite, the delta peaks in $A_n(\omega)$ and $A_m(\omega+\omega_1)$ obtain a finite width. Due to this broadening, the gap effectively becomes smaller than it would be for an infinitesimally small $\eta$, resulting in an increase of the GQM because of the factor $1/\omega_1$ in Eq.~(\ref{defGQM}). To confirm the validity of the above discussion, we investigated the $\eta$ dependence of the GQM in Fig.~\ref{fig:Fig5}. This figure shows that the general behavior of the GQM follows the behavior of the existing definition in the noninteracting system for any $\eta$. Figure~\ref{fig:Fig5} shows that the GQM becomes large when $\eta$ is increased. For any $\eta$, the difference between the existing quantum metric and the GQM becomes maximal at the $\mathrm{M}$ point of the BZ, where the gap is the smallest. 
Because $\eta$ can be regarded as a scattering rate due to defects or impurities(\cite{PhysRevB.105.085154, PhysRevB.105.085139, PhysRevB.102.165151}), our result implies that the GQM will increase in the presence of the defects or impurities.

\begin{figure}[H]
\centering
\includegraphics[width=\linewidth]{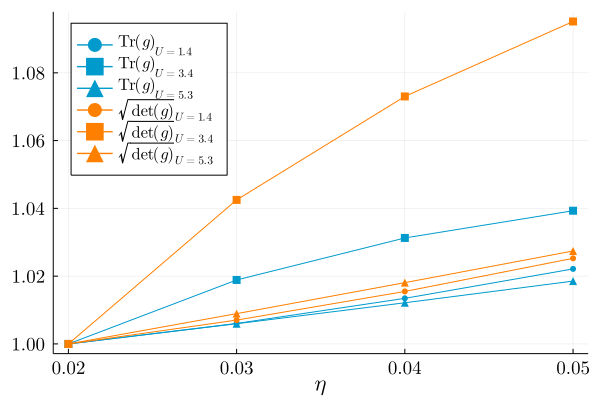}
\caption{Dependence of $\trg$ and $\detg$ on $\eta$ for different interaction strengths. The points in the figure represent the value of the integrals divided by $\trg$ at $\eta=0.02$ and $\detg$ at $\eta=0.02$}
\label{fig:etadep}
\end{figure}

Finally, in Fig.~\ref{fig:etadep}, the $\eta$ dependence of the integral of the GQM is shown for different interaction strengths. We see that the integral increases with increasing $\eta$ and  $\detg$ at $U=3.4$, nearest to the topological phase transition, is the most sensitive to $\eta$. These results suggest that a small peak of $\detg$ in Fig.~\ref{fig:Wilson_M=1.5}(bottom) and Fig.~\ref{fig:WilsonU=2.2}(bottom) are numerical artifacts created by the finite $\eta$.

\bibliography{reference.bib}

\end{document}